\documentclass{andp2012}
\usepackage[english]{babel}
\usepackage{amsmath}
\usepackage{amssymb}
\usepackage{graphicx}
\usepackage{grffile} 
\usepackage{float}
\usepackage{color}
\usepackage{subfigure}
\usepackage{multicol}

\newcommand{\kk}{\mathbf{k}}
\newcommand{\qq}{\mathbf{q}}
\newcommand{\A}{\mathbf{A}}

\newcommand{\trel}{t_\mathrm{rel}}
\newcommand{\tave}{t_\mathrm{ave}}

\newcommand{\tmin}{t_\mathrm{min}}
\newcommand{\resigma}{\mathrm{Re}\ \Sigma}
\newcommand{\imsigma}{\mathrm{Im}\ \Sigma}
\newcommand{\BSCCO}{Bi$_{2}$Sr$_{2}$CaCu$_{2}$O$_{8+x}$}

\author[kemper]{A.F. Kemper\inst{1}}
\author[sentef]{M.A. Sentef\inst{2}}
\author[moritz]{B. Moritz\inst{3}}
\author[devereaux]{T. P. Devereaux\inst{3,4}}
\author[freericks]{J.K. Freericks\inst{5}\footnote{Corresponding author\quad E-mail:~\textsf{james.freericks@georgetown.edu}}}
\address[1]{Department of Physics, North Carolina State University, Raleigh, NC  27695, USA}
\address[2]{Max Planck Institute for the Structure and Dynamics of Matter, Center for Free Electron Laser Science, 22761 Hamburg, Germany}
\address[3]{Stanford Institute for Materials and Energy Sciences (SIMES), SLAC National Accelerator Laboratory, Menlo Park, CA 94025, USA}
\address[4]{Geballe Laboratory for Advanced Materials, Stanford University, Stanford, CA 94305, USA}
\address[5]{Department of Physics, Georgetown University, Washington, DC 20057, USA}

\shortauthors{A. F. Kemper et al.}

\title[Review of the theoretical description of time-resolved angle-resolved photoemission spectroscopy $\cdots$]{Review of the theoretical description of time-resolved angle-resolved photoemission spectroscopy  in
electron-phonon mediated superconductors}

\begin{abstract}
We review recent work on the theory for pump/probe photoemission spectroscopy of electron-phonon mediated superconductors in both the normal and the superconducting states. We describe the formal developments that allow one to solve the Migdal-Eliashberg theory in nonequilibrium for an ultrashort
laser pumping field, and explore the solutions which illustrate the relaxation as energy is transferred from electrons to phonons. We focus on exact results emanating from sum rules and approximate numerical results which describe rules of thumb for relaxation processes. In addition, in the superconducting state, we describe how Higg's oscillations can be excited due to the nonlinear coupling with the electric field and
how pumping the system can enhance superconductivity.
\end{abstract}

\begin{document}

\maketitle


\section{Introduction}
The recent availability of ultrashort and ultraintense photon sources ranging from conventional lasers to free-electron lasers and encompassing a wide range of the electromagnetic spectrum has resulted in a blossoming of experiments in pump/probe studies of nonequilibrium phenomena in solids. In these experiments, an intense ultrashort electric field pulse excites the system into nonequilibrium, which is later probed by a second (typically weaker) pulse with some time delay with respect to the pump pulse. While reflectivity experiments have been examined for decades, recently many new experiments ranging from X-ray diffraction to photoemission have been explored. In this review, we focus on a summary of time-resolved photoemission studies on electron-phonon mediated systems. Due to the length restrictions of this review, we focus primarily on the work of our group. Additional work has been completed by the Eckstein, Oka and Werner groups, with a focus on different aspects of the problem than what we discuss here, see Ref.~\cite{aoki_nonequilibrium_2014} for an extensive review, and
other recent work~\cite{werner_eckstein_2013,werner_eckstein_2015,murakami_etal_2015}.

Dynamical mean-field theory (DMFT)  is now considered one of the most accurate approaches for solving the many-body problem. In 2006, it was generalized from equilibrium problems to nonequilibrium~\cite{freericks_nonequilibrium_2006}, and since then has been used to examine a wide range of different nonequilibrium systems. In this work, we focus on Migdal-Eliashberg theory to describe electron-phonon coupled systems~\cite{migdal_1958,eliashberg_1960,eliashberg_1961}. While the original Migdal-Eliashberg theory predates DMFT, it also works with a local self-energy, and can be interpreted as the first application of DMFT. Generalized to nonequilibrium, it continues to be able to be interpreted in this DMFT context, but with differing forms of self-consistency depending on how the problem is formulated and how the solution is carried out. We describe these subtle issues in detail below.

\section{Method}

\subsection{Model}
We primarily study the Hubbard-Holstein model here~\cite{hubbard_1963,holstein_1959}
\begin{align}
\begin{split}
\mathcal H=& \sum_{\kk,\sigma} \epsilon(\kk) c^\dagger_{\kk\sigma} c_{\kk\sigma}^{\phantom\dagger} + \sum_i U n_{i\uparrow} n_{i\downarrow}
\nonumber\\
		 &+\sum_{\qq,\gamma} \Omega_{\qq,\gamma}  b_{\qq,\gamma}^\dagger b_{\qq,\gamma}^{\phantom\dagger} - 
		\sum_{\qq,\gamma,\sigma} g_\gamma  c_{\kk+\qq,\sigma}^\dagger c_{\kk,\sigma}^{\phantom\dagger} \left( b_{\qq,\gamma}^{\phantom\dagger} + b_{-\qq,\gamma}^\dagger \right)  ,
\end{split}
\end{align}
where the individual terms consecutively represent the kinetic energy of the electrons with a band-structure $\epsilon(\kk)$, 
the on-site electron-electron repulsion $U$,
the total energy Einstein phonons in branches $\gamma$ with a frequency $\Omega_{\qq,\gamma}$, and an electron-phonon coupling term of strength $g_\gamma$. Typically, we will
integrate over the phonon momenta and work with the phonon distribution function $\alpha^2 F(\Omega)$.
Furthermore, $c^\dagger_\alpha (c_\alpha^{\phantom\dagger})$ are the standard creation (annihilation) operators for an electron in state $\alpha$ (where $\alpha$ denotes momentum and spin or lattice site and spin, as determined by the context of the given operator); similarly, $b^\dagger_{\qq,\gamma} (b_{\qq,\gamma}^{\phantom\dagger})$ creates (annihilates) a phonon with momentum $\qq$ in branch $\gamma$.  The electron-phonon coupling is the conventional coupling between the electron charge and the phonon coordinate; for a harmonic oscillator, this coupling is identical to that of the fluctuations of the charge coupling to the phonon coordinate, since the two are related simply by a shift of the origin of the phonon coordinate.
For concreteness we study this model on a square lattice with a band-structure given by nearest neighbor hopping ($t_{nn})$,
\begin{align}
\epsilon(\kk) = &-2 t_{nn} \left[ \cos(k_x) + \cos(k_y)\right] -\mu
\end{align}
where $\mu$ is the chemical potential.  We have used the convention that $\hbar=c=e=1$, which makes the unit of time to be given by the inverse energy.  

Within a DMFT approach, we have to solve an impurity problem associated with the lattice. This
impurity problem can be solved in many different ways. Here, we choose to invoke perturbation
theory as the solver of choice for the impurity problem. This allows us to examine a range of different problems in the weak-coupling realm (and even into the intermediate-coupling realm when 
vertex corrections are small for the electron-phonon self-energy). The issue that always arises involves the level of self-consistency imposed. In the equilibrium theory for electron-phonon mediated superconductors, we work with the fully dressed phonons, extracted from experiment, and hence we do not renormalize the phonons at all, only the electrons are dressed self-consistently by the phonons. Such an approach might begin to fail in nonequilibrium if a significant amount of energy is transferred to the phonon bath, because it will heat up and can change its properties. Nevertheless, we continue to treat the phonons as fully dressed phonons, even in nonequilibrium here, which is accurate for short times, and likely to be a reasonably good approximation except for strong or resonant THz pumping.

Hence, the electron-phonon interaction part of the self-energy is treated at the self-consistent Born level (or second-order perturbation theory with respect to $g$, performed self-consistently for the electrons)
where the self-energy satisfies
\begin{align}
\bar\Sigma^c_{el-ph}(t,t') = i g^2 \bar\tau_3\ \bar G_\mathrm{loc}^c(t,t') \bar\tau_3\ D^c_0(t,t'),
\label{eq: se_elph}
\end{align}
where $\bar\tau_3$ is the $z$ Pauli matrix in Nambu space,
and $\bar G^c_\mathrm{loc}(t,t') = N_\kk^{-1} \sum_\kk \bar G^c_\kk(t,t')$ 
i.e. the nonequilibrium, two-time, contour-ordered local Green's function. Note that the phonon propagator remains the bare propagator in the theory. Multiple phonon modes are taken into
account via
\begin{align}
D^c_0(t,t') = \int d\Omega \alpha^2F(\Omega) D_0^c(t,t';\Omega).
\end{align}
$D_0^c(t,t';\Omega)$ is the bare propagator for a phonon with a single mode frequency $\Omega$
\cite{mahan}:
\begin{align}
D_0^c(t,t';\Omega) =& -i\left[ n_B(\Omega/T) + 1- \theta^c(t,t')\right] e^{i\Omega(t-t')} \nonumber \\
& -i\left[ n_B(\Omega/T) + \theta^c(t,t')\right] e^{-i\Omega(t-t')},
\end{align}
where $n_B(x)$ is the Bose distribution function, and $\theta^c(t,t')$ is the contour-ordered
Heaviside function.

The electron-electron
scattering part of the self-energy is evaluated at the level of a self-consistent second-order perturbation theory in $U$,
\begin{align}
\bar\Sigma^c_{el-el}(t,t') =& U^2 \bar\tau_3\ \bar G_\mathrm{loc}^c(t,t') \bar\tau_3\ 
\times \nonumber\\
&\mathrm{Tr}\left\{ \bar G_\mathrm{loc}^c(t,t') \bar\tau_3 \bar G_\mathrm{loc}^c(t',t) \bar\tau_3 \right\}.
\end{align}
This a second-order conserving approximation, as opposed to an iterated perturbation theory approximation.

\begin{figure}
	\includegraphics*[width=0.9\columnwidth]{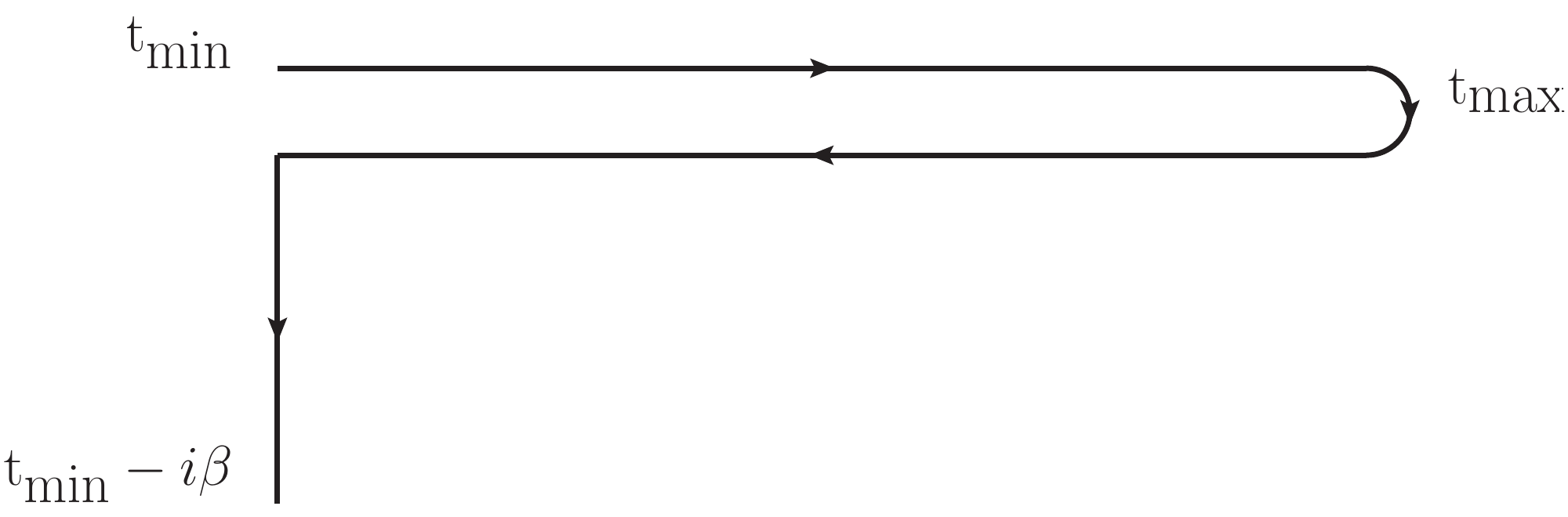}
	\caption{The Kadanoff-Baym-Keldysh contour~\cite{kadanoff_baym_1963,keldysh_1965} used in the calculations. It starts from an initial time
$t_{\rm min}$, runs to a maximum time $t_{\rm max}$, returns back to the initial time, and then runs parallel to the imaginary axis a length $\beta$ given by the inverse of the initial equilibrium temperature of the system before it is pumped $t_{\rm min}-i\beta$.}
	\label{fig:k_contour}
\end{figure}

We utilize the standard two-time Keldysh formalism. The
contour-order Green's functions (denoted with a $c$ superscript) are either $1\times 1$ or $2\times 2$ matrices in Nambu space\cite{AGD,rammer}, depending on whether the solution to an ordered state is being sought. For the superconducting case, where we have
$2 \times 2$ matrices, we find
\begin{align}
\bar{G}^c_\kk(t,t') &= -i\left\langle\mathcal{T}_c
\left(
\begin{array}{cc}
c_{\kk\uparrow}^{\phantom\dagger}(t) c^\dagger_{\kk\uparrow}(t') & c_{\kk\uparrow}^{\phantom\dagger}(t) c_{-\kk\downarrow}^{\phantom\dagger}(t') \\
c^\dagger_{-\kk\downarrow}(t) c^\dagger_{\kk\uparrow}(t')  & c^\dagger_{-\kk\downarrow}(t)  c_{-\kk\downarrow}^{\phantom\dagger}(t')
\end{array}
\right)\right\rangle \\
&\equiv
\left(
\begin{array}{cc}
G^c_\kk(t,t') & F^c_\kk(t,t') \\ 
F^{\dagger c}_\kk(t,t')  & -G^c_{-\kk}(t',t) 
\end{array}
\right).
\end{align}
Here, the angle brackets denote a trace over all states weighted by the initial equilibrium density matrix at an initial temperature $T$: $\rho(T)=\exp(-\beta\mathcal{H})/\mathcal{Z}$ with $\beta=1/T$ and $\mathcal{Z}={\rm Tr}\exp(-\beta\mathcal{H})$ is the partition function (we set $k_B=1$).
In the normal state, the off-diagonal elements are zero and the two remaining
components are redundant: thus only the $(1,1)$ component is kept.
Here, $t$ and $t'$ lie on the Keldysh contour, and $\mathcal{T}_c$ denotes time-ordering along the Kadanoff-Baym-Keldysh contour.

\subsection{Numerical approach}

We solve the equations of motion (on the contour, 
shown in Fig.~\ref{fig:k_contour}):
\begin{align}
\begin{split}
\left(i\partial_t \bar\tau_0 - \bar\epsilon_\kk(t) \right) \bar{G}^c_\kk(t,t')
&= \delta^c(t,t') \bar\tau_0 
\\&+ \int_c d\bar t\  \bar\Sigma^c(t,\bar t) \bar G_\kk^c(\bar t,t')
\end{split}
\end{align}
with the Nambu bandstructure given by the Peierls' substitution
\begin{align}
\bar\epsilon_\kk(t) &= \left(
\begin{array}{cc}
\epsilon_\uparrow(\kk-\A(t)) & 0 \\
0 & -\epsilon_\downarrow(-\kk-\A(t))
\end{array}
\right)
\end{align}
where 
$\bar\tau_0$ is the identity matrix, $\epsilon_\uparrow(\kk) = \epsilon_\downarrow(\kk) = \epsilon(\kk)$ is the
bare bandstructure for a spin up/down electron, and $\A(t)$ is the vector potential in the Hamiltonian gauge.

The contour equation of motion can be separated into separate Matsubara ($M$), lesser ($<$), and greater ($>$) Green's functions, 
as well as the mixed real-imaginary $\rceil$ and $\lceil$ types.
These each have an equation of motion, which we list here for completeness.
The equations are solved using a large-scale parallel computational approach, as described
in Ref.~\cite{stefanucci_book}.

\begin{subequations}
\begin{align}
\big[ -\partial_\tau \bar\tau_0 - \bar\epsilon_\kk(\tmin)\big]& \bar G_\kk^M(\tau) = i\delta(\tau)\bar\tau_0 \nonumber\\
-& i \int_0^\beta d\bar \tau  \bar\Sigma^M(\tau-\bar\tau) \bar G_\kk^M(\bar\tau), \\
\big[ i\partial_t \bar\tau_0- \bar\epsilon_\kk(t) \big] &G_\kk^\rceil(t,-i\tau) = \nonumber\\
&\int_{\tmin}^t d\bar t\ \bar \Sigma^R(t,\bar t) \bar G_\kk^\rceil(\bar t, -i\tau)  \nonumber\\
-&i \int_0^\beta d\bar\tau\ \bar\Sigma^\rceil(t,-i\bar\tau) \bar G_\kk^M(\bar\tau -\tau),\\
\big[ i\partial_t \bar\tau_0- \bar\epsilon_\kk(t) \big]& \bar G_\kk^\gtrless(t,t') =\nonumber\\ 
&\int_{\tmin}^t d\bar t\  \bar\Sigma^R(t,\bar t) \bar G_\kk^\gtrless(\bar t, t') \nonumber\\
+&\int_{\tmin}^{t'} d\bar t\  \bar\Sigma^\gtrless(t, \bar t) \bar G_\kk^A(\bar t, t') \nonumber\\
 -& i \int_0^\beta d\bar\tau\ \bar\Sigma^\rceil(t,-i\bar\tau) \bar G_\kk^\lceil(-i\bar\tau,t'),
 \label{eq:Gless}
\end{align}
\label{eq:eoms}
\end{subequations}
Here,we typically use $t_\mathrm{min}=0$ 
without loss of generality. Once the full set of time-dependent equations are solved self-consistently, we have the time-dependent Green's functions and self-energies, which are employed to calculate observables.

The derivation of the tr-ARPES spectra is complicated, in general, but simplifies when one restricts to a single-band model and employs the constant matrix element approximation. In this case, the tr-ARPES spectra
can be computed from the probe-pulse-weighted relative-time Fourier transform of the occupied (lesser) Green's function\cite{freericks_09}
\begin{align}
I(\kk,\omega,t_0) =\mathrm{Im} \int dt dt' 
p(t,t',t_0)
e^{i\omega(t-t')} G_{\tilde\kk(t,t')}^<(t,t').
\label{eq:trarpes}
\end{align}
Here, $p(t,t',t_0)$ denotes a two-dimensional Gaussian probe with a probe width of $\sigma_p$ centered
at $(t,t')=(t_0,t_0)$. The shift in the momentum $\kk$ due to the vector potential has to be corrected
via a gauge shift in $G^<_\kk$ as\cite{v_turkowski_book}
\begin{align}
\tilde\kk(t,t') = \kk + \frac{1}{t-t'} \int_{t'}^{t} d\bar t\, \A(\bar t).
\end{align}

The current is computed from the density as
\begin{align}
\mathbf{j}(t) = N_\kk^{-1} \sum_\kk \mathbf\nabla \epsilon(\kk-\A(t))\ \mathrm{Im}\ G_\kk^<(t,t)
\end{align}
where the derivative is taken along the field $(11)$ direction.  Even in the superconducting
phase, the (1,1) component of the Nambu matrix is used \textemdash\ this gives the supercurrent
as well as the normal current.

\section{Time-resolved dynamics of the normal state}

When Green's functions were first introduced into many-body physics, it was rapidly recognized that the relaxation time to a perturbation of the retarded Green's function was given by the inverse of the imaginary part of the self-energy evaluated at the pole of the retarded Green's function that lay closest to the real axis but below it~\cite{galitskii1958translation}. For systems described by Fermi liquids, this was well approximated by the equilibrium self-energy evaluated at the given frequency. Since then, it has been generally believed that the imaginary part of the self energy will continue to govern the relaxation processes, even in nonequilibrium. In fact, for the simplest version of an electron-phonon coupled system, one can prove that this is the case~\cite{sentef_13,kemper_relationship_16}. 

It has recently come to light, that the situation in nonequilibrium often is different from this simple behavior. This arises, from a mathematical standpoint, from the fact that one needs to examine the evolution of the electron population $\langle n_\kk\rangle$ as a function of time, which necessarily brings in the average time dependence of the lesser self-energy, which may not behave the same way as the relative time dependence of the retarded self-energy~\cite{kemper_relationship_16,kemper_role_16}. Experimentally, this has already been seen clearly, as the decay of populations is governed by different time scales than the widths of ARPES peaks in equilibrium\cite{yang_inequivalence_2015}. That work shows that there is a marked contrast between
the lifetime of a singly-excited electron and the decay rates of the population. As noted above,
these quantities in principle arise from orthogonal directions in the time domain \textemdash
along $\tave$ and along $\trel$.  We shall illustrate this by first focusing on the simple
case of electron-phonon (el-ph) coupling for an Einstein mode with energy $\Omega$,
and then including electron-electron (el-el) interactions.

\subsection{Electron-phonon interactions}

\begin{figure}
	\includegraphics[width=0.99\columnwidth]{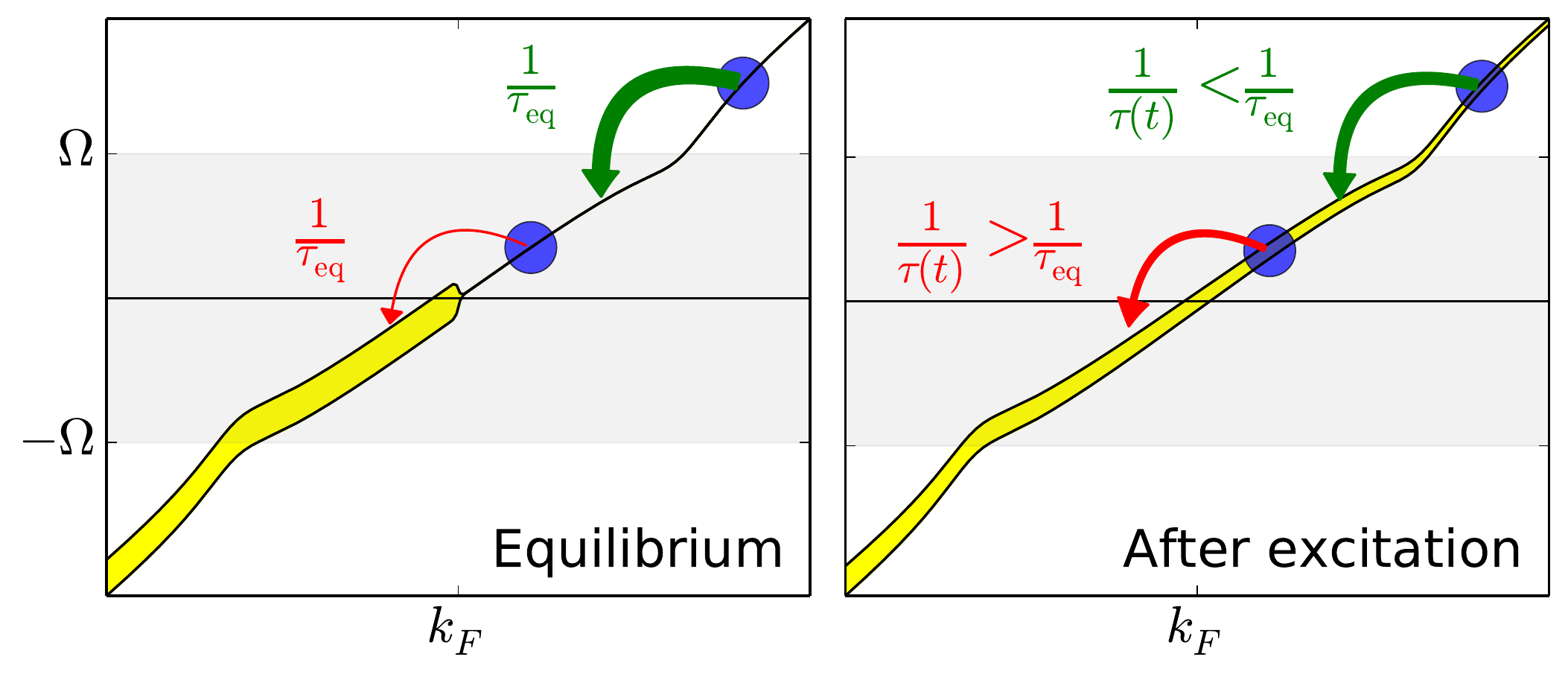}
	\caption{Phase space restrictions on the scattering of a single excited quasiparticle.
	in equilibrium (left) and after the excitation by a pump laser
	pulse (right). Figure reprinted with permission from~\cite{kemper_effect_2014} \copyright ~American Physical Society.}
	\label{fig:phase_space}
\end{figure}

\begin{figure}
	\includegraphics[width=0.99\columnwidth,clip=true, trim=0 850 0 0]{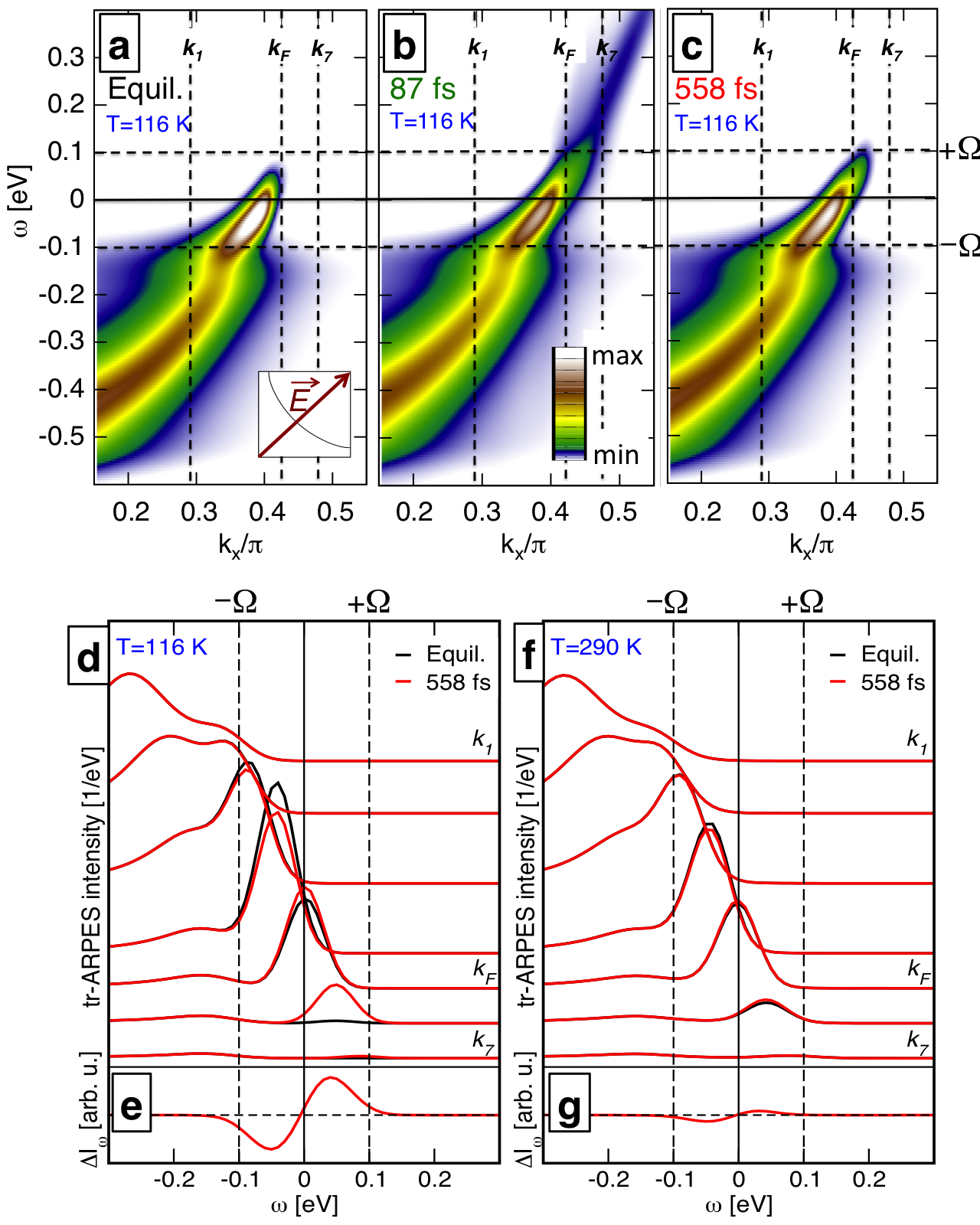}
	\caption{tr-ARPES spectra along the zone diagonal in the normal state with
	a single phonon mode at $\Omega=0.1$ eV. The panels are at various times:
	in equilibrium (a), just after the pump (b) and long after the pump (c). Reprinted with permission from~\cite{sentef_examining_2013} \copyright ~American Physical Society.}
	\label{fig:PRX_fig1}
\end{figure}

Let us begin by discussing the origin of the quasiparticle lifetime in equilibrium, as given by a phase space argument. The electron-phonon coupling involves an inelastic scattering, because phonons are created or destroyed in the interaction, and they carry an energy $\Omega$. Hence, electrons cannot be scattered at low-energies, but must have an energy larger than $\Omega$ in order to create a phonon and scatter. This further implies that quasiparticles that are excited to energies within the phonon energy
$\Omega$ above the Fermi level $E_F$ (which we call the ``phonon window'')
are Pauli blocked from further scattering, and thus have a long
lifetime (narrow line width). Quasiparticles with energies above $\Omega$ have no such restriction,
and thus have a short lifetime (high scattering rate and wide line width). This restriction is
illustrated schematically in Fig.~\ref{fig:phase_space}.  In experimental spectra, this leads to a sharp step in
the line width due to the sudden increase in the scattering rate [$\imsigma(\omega)$] at the
phonon energy\cite{zhou_handbook_2007}, as shown in Fig.~\ref{fig:PRX_fig1} The sharp step
in $\imsigma(\omega)$ occurs jointly with a peak in $\resigma(\omega)$ through the
Kramers-Kronig relation, leading to a kink in the ARPES spectrum at the phonon energy $\Omega$,
also seen in the experimental spectrum.

\begin{figure}
	\includegraphics[width=\columnwidth]{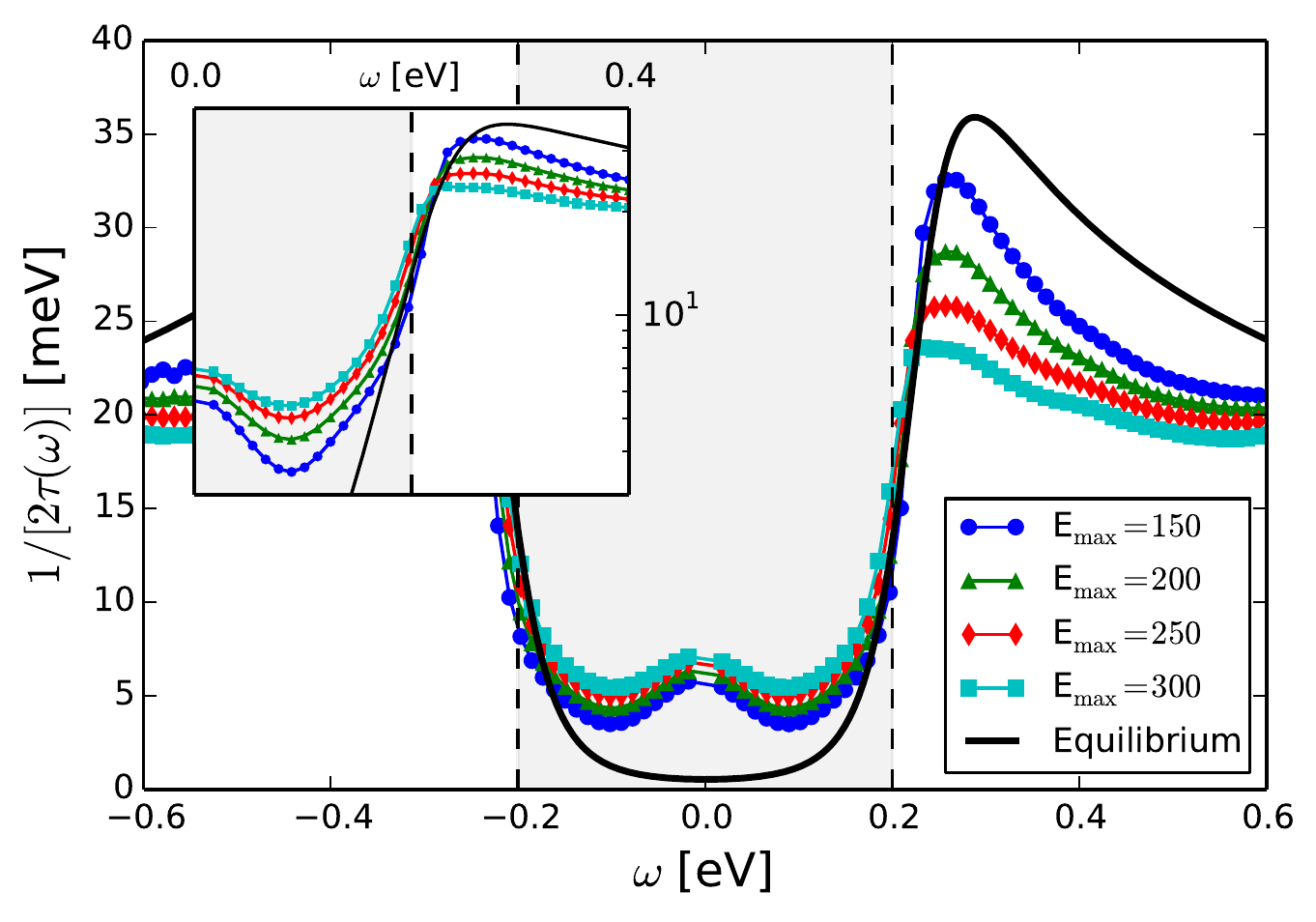}
	\caption{Population decay rates obtained from single-exponential fits. Reprinted with permission from~\cite{kemper_effect_2014} \copyright ~American Physical Society.}
	\label{fig:PRB_onebytau}
\end{figure}

The marked difference in scattering rates is also observed in the population dynamics for this
electron-phonon system, although with some differences that are unique to the nonequilibrium
process. After excitation with the pump laser, the populations are measured and their return to
equilibrium is characterized by some single- or multi-exponential curve
\cite{kampfrath_strongly_2005,perfetti_ultrafast_2007,kirchmann_quasiparticle_2010,graf_nodal_2011,torchinsky_nonequilibrium_2011,cortes_momentum-resolved_2011,rettig_ultrafast_2012,smallwood_tracking_2012,ulstrup_ultrafast_2015,yang_inequivalence_2015}.
Typical spectra during and after the pump are shown in Fig.~\ref{fig:PRX_fig1}.
In the limit of zero pump fluence, the population decay rates [$1/\tau(\omega)$] can be shown
to approach the quasiparticle scattering rates\cite{kemper_mapping_2013,sentef_examining_2013}
as plotted in Fig.~\ref{fig:PRB_onebytau}.
However, the pump also modifies the distribution of electrons, which in turn affects the interactions:
the self-energy here is given by the normal-state limit of Eq.~(\ref{eq: se_elph}):
\begin{align}
\Sigma^c_{el-ph}(t,t') = \frac{i g^2}{N_\kk} \sum_\kk G^c_\kk(t,t') D_0^c(t,t'),
\end{align}
where $D_0^c(t,t')$ is the bare phonon propagator defined above.
Thus, the interactions ``know'' about the distribution of the electrons through $G^<_\kk(t,t')$.
This is reflected in the simple phase space picture through a redistribution of the spectral weight, 
leading to a modification of the scattering rates. Within the phonon window, the scattering rate
increases since a larger phase space for scattering becomes available; outside the phonon
window, the opposite occurs.  This is also observed in the population dynamics \textemdash
$1/\tau(\omega)$ obtained from the population decay shows a fluence dependence that agrees with
the simple picture of phase-space restriction, but with a modification due to the pump field
(see Fig.~\ref{fig:PRB_onebytau}).  This effect has been observed experimentally
in the tr-ARPES spectra of \BSCCO~\cite{rameau_energy_2015}.

\begin{figure}
	\includegraphics[width=\columnwidth,clip=true,trim=0 290 0 0]{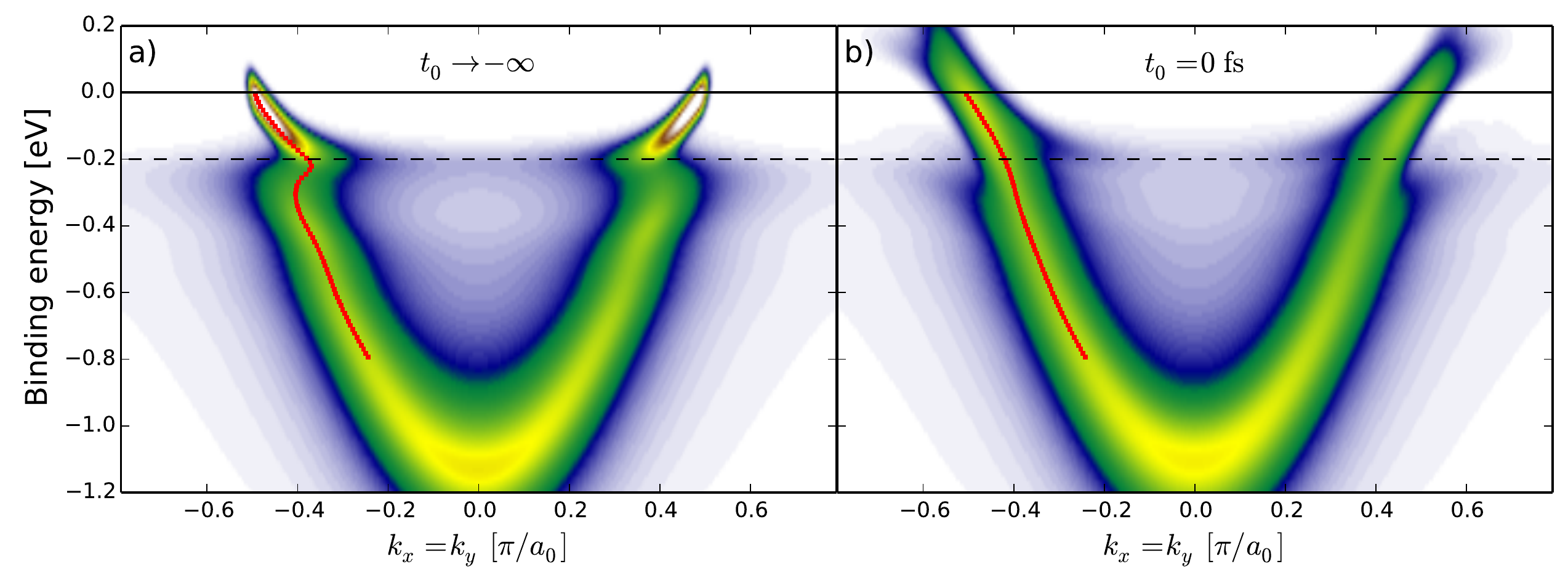}
	\caption{Weakening of the electron-phonon kink due to the pump. Reprinted with permission  from~\cite{kemper_effect_2014} \copyright ~American Physical Society.}
	\label{fig:PRB_kinksoftening}
\end{figure}

In addition to the softening of the step in the lifetimes [$\imsigma(\omega)$], the sharp peak in
$\resigma(\omega)$ is reduced, leading to an apparent weakening of the kink at $\omega=\Omega$
as shown in Fig.~\ref{fig:PRB_kinksoftening}, and observed experimentally by several groups
\cite{zhang_ultrafast_2014, rameau_14,ishida_quasi-particles_2016}.
Because the kink is historically used as a quantitative measure of the electron-phonon coupling strength
in equilibrium, this softening was initially ascribed to a dynamical decoupling of electrons from the phonons in nonequilibrium. While it is true that a weakening of the electron-phonon coupling strength would reduce the kink, and one might argue such a weakening is possible due to how the nonequilibrium electrons screen differently than the original equilibrium electrons, one must contrast that reasoning with an exact analysis of what happens in the Holstein-Hubbard Hamiltonian. In other words, there is another way that the kink can be softened, and that is via a redistribution of spectral weight (and of electron populations) which reduce the phase-space limitations of equilibrium. 
Indeed, it was shown\cite{kemper_effect_2014, freericks_nonequilibrium_2014} that there is a sum rule
for the self-energy, and that the zeroth moment of the imaginary part of the self-energy is preserved out of equilibrium
as long as the phonon fluctuations are not changed.  The perturbative form of the sum rule reads as follows in the time domain:
\begin{align}
{\rm Im}\Sigma^R_{el-ph}(t,t) = -g^2 \left[ 2 n\left (\frac{\Omega}{T}\right ) + 1 \right],
\end{align}
where $n(\Omega/T)$ is the Bose function evaluated at the phonon frequency, while the exact relation, which holds for all cases, reads:
\begin{align}
{\rm Im}\Sigma^R_{el-ph}(t,t)=-g^2\left [ \langle x_i^2(t)\rangle-\langle x_i(t)\rangle^2\right ],\end{align}
where $x_i(t)$ is the operator for the phonon coordinate at lattice site $i$ in the Heisenberg representation (and is independent of $i$ in a homogeneous system). Hence, the theoretical results for the Holstein-Hubbard model directly show the kink softening resulting from a redistribution of the spectral weight (or equivalently the phase space for electron-electron scattering) with the electron-phonon coupling (as measured by the zeroth moment of the retarded self-energy)
remaining constant, independent of the fluence, until it causes a change in the phonon fluctuations
as a function of time. Since the phonons are not renormalized in the current calculations, this latter effect never occurs.

\subsection{Electron-electron interactions}
The inclusion of other types of interactions reveals an even further stark difference between
equilibrium lifetimes of a single excited quasiparticle and time-resolved population dynamics
\cite{yang_inequivalence_2015}. The issue at hand is that there is a fundamental difference between these
quantities. A singly excited quasiparticle has a lifetime due to its wave function decay, which can
involve scattering into another state, or simply dephasing due to e.g. impurity scattering.
On the other hand, a macroscopic population that has been excited due to a laser pulse has
absorbed energy and thus can only return to equilibrium if it releases said energy. In pump-probe
experiments, this can happen either through the coupling of the electron population to some
bath \textemdash the phonons \textemdash or through diffusion of the excited population
away from the excitation volume. Here, we focus on the former. Electron-phonon interactions
carry energy from the electrons into the phonon bath, and thus provide a path to return to equilibrium.

Electron-electron interactions, on the other hand, maintain the energy within the electronic subsystem.
While the electrons can individually exchange energy, their total energy remains fixed; this is an exact statement, true for any isolated system after the pump is turned off. This limits
the action of el-el interactions to simply causing a quasithermalization at some elevated effective temperature,
after which the new high temperature equilibrium is maintained. This is where the relation
between the equilibrium self-energy and the population dynamics completely breaks down \textemdash the
self energy reflecting the el-el interactions is still present, albeit at a higher temperature,
yet the population is not returning to the original equilibrium.

\begin{figure}
	\includegraphics[width=0.8\columnwidth,clip=true,trim=0 0 0 0]{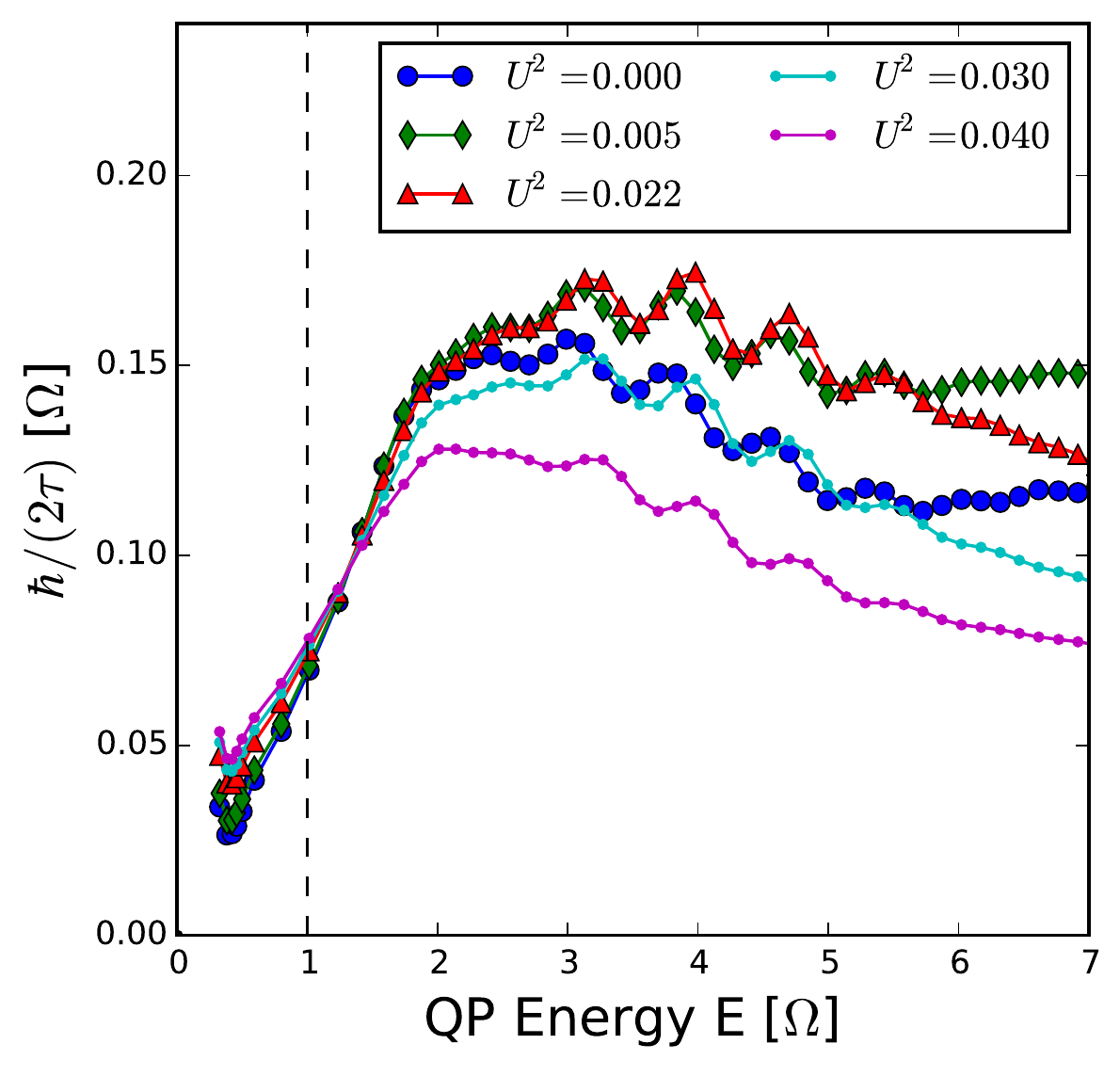}
	\caption{Decay rates of a system with el-el and el-ph scattering for various values of
	the el-el coupling strength $U$ (reproduced from Ref.~\cite{rameau_energy_2015}).}
	\label{fig:bi2212_onebytau}
\end{figure}

In real systems, both el-el and el-ph scattering are present. This combination causes a complex
dynamics where the two interactions each push towards its individual final state \textemdash
equilibration at the current energy within the electrons (for el-el) and equilibration with the phonons
(for el-ph).  Since el-ph scattering is responsible for carrying energy out of the electronic subsystem, and since it does so in finite quanta,
the phase space restrictions discussed above leave an imprint on the population decay rates.
This can be seen in Fig.~\ref{fig:bi2212_onebytau}, where the decay rates are
shown for various strengths of el-el scattering.
The step in the decay rates at the quasiparticle energy $E=\Omega$ seen in
the case with only el-ph coupling remains even as the el-el interactions
are turned on, even to the point where the total interaction energy for el-el scattering
outweighs the el-ph scattering. This prediction was recently confirmed in time-resolved
experiments on \BSCCO~\cite{rameau_energy_2015}.

\begin{figure*}[htpb]
\begin{centering}
	\includegraphics*[width=0.75\textwidth]{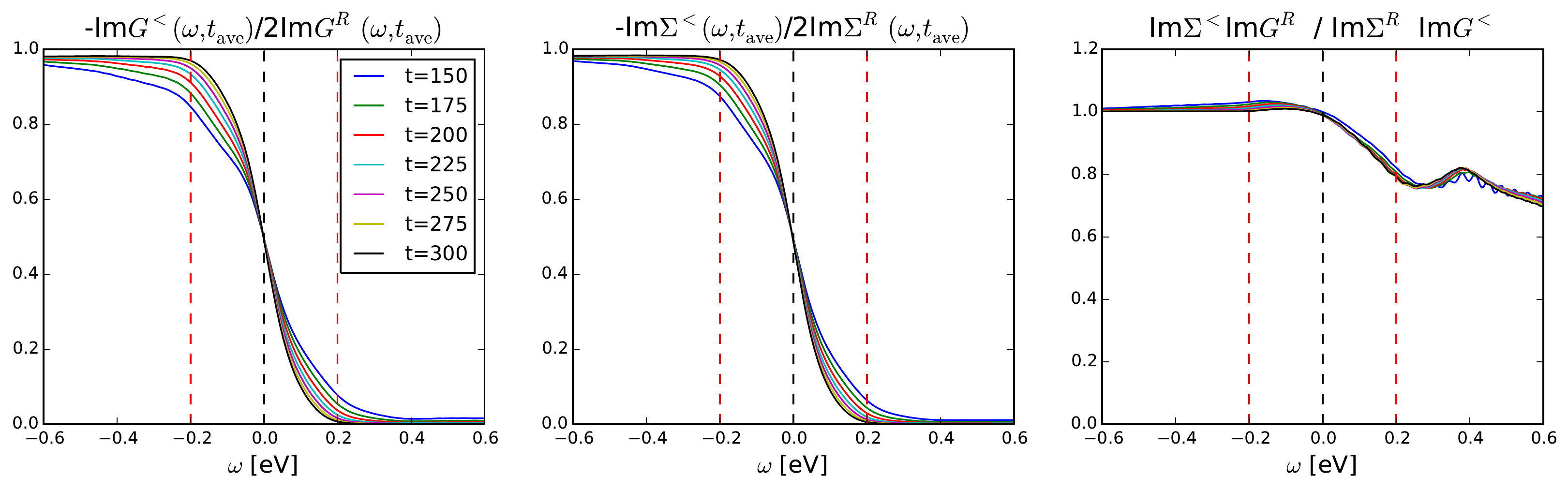}
	\caption{Plot of the effective distribution functions $f_G$ (left) and $f_\Sigma$ (center)
	and their ratio (right) for the local Green's function and self-energy. Reprinted with permission from
\cite{kemper_role_16} }
	\label{fig:fortschritte}
\end{centering}
\end{figure*}

Finally, it is important to discuss the question of the applicability of a hot-electron model. An exact analysis of the equation of motion for electron populations shows that once one uses a single 
distribution function for the lesser Green's function and the lesser self-energy, then the population no longer evolves with time~\cite{kemper_relationship_16}. The distribution doesn't even need to be an equilibrium one. This immediately shows that a hot-electron model can never be employed exactly in describing population dynamics, because it has the distribution function for equilibrium, but with a time-dependent temperature. Unfortunately, the exact dynamics precludes such behavior. Instead, the distribution functions for the Green's function $f_G$ and the self-energy $f_\Sigma$ both evolve distinctly from one another, even though they remain close. This is shown in Fig.~\ref{fig:fortschritte}, where we plot the distribution function for the lesser Green's function, for the lesser self-energy and their ratio for an electron-phonon coupled  system that is relaxing in the long-time limit after the pump has been applied~\cite{kemper_role_16}.

\begin{figure}[htpb]
	\includegraphics[width=0.99\columnwidth]{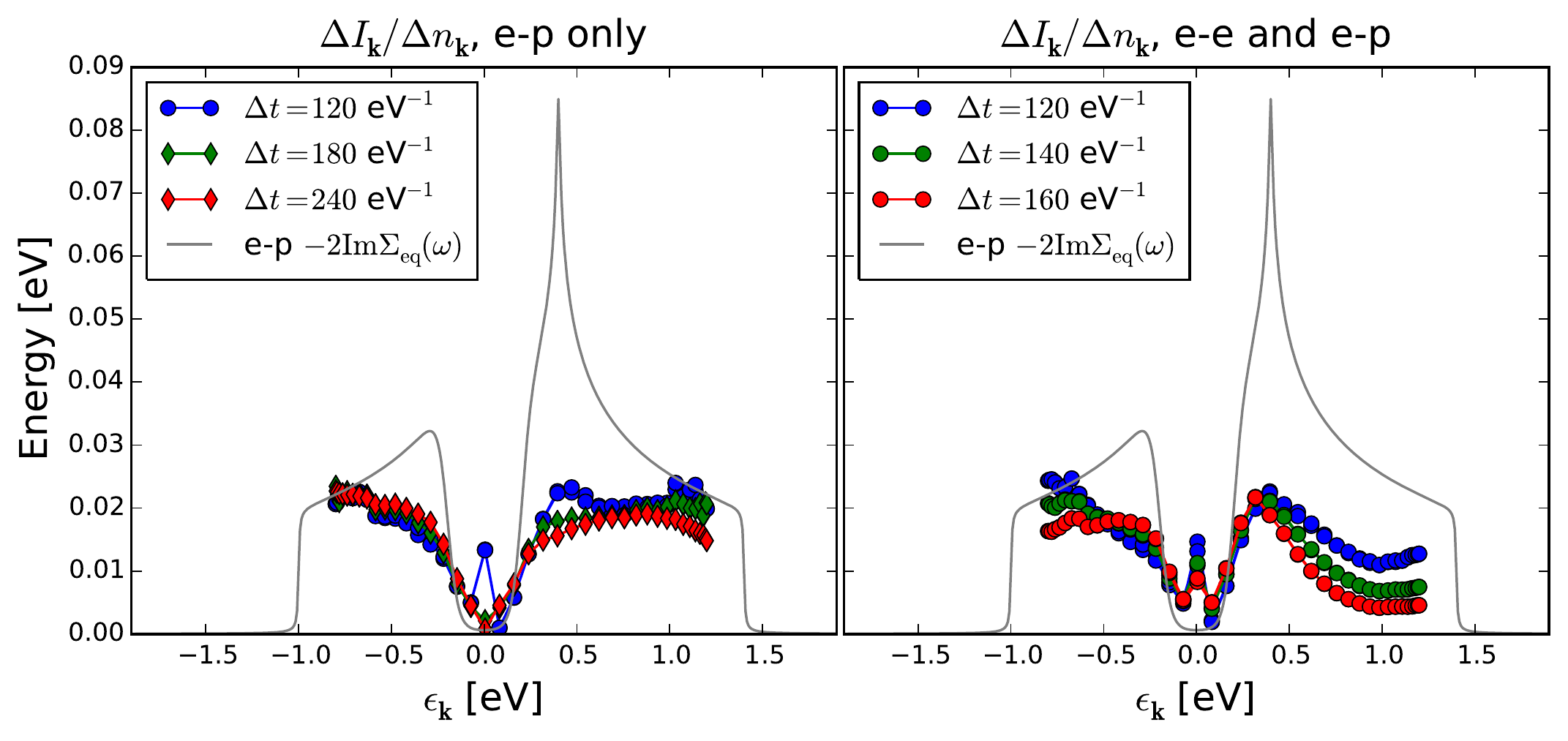}
	\caption{Approximate relaxation rate extracted from the change in scattering integrals:
	$\Delta I_\kk/ \Delta n_\kk$ after a pump has been applied.The equilibrium electron-phonon self-energy is shown for reference. Reprinted with permission from~\cite{kemper_relationship_16}.}
	\label{fig:entropy_fig4}
\end{figure}

So, given this complicated behavior in nonequilibrium, one may ask just what is it that does determine the relaxation rate? It turns out, empirically, that the result is close to the imaginary part of the self-energy, but somewhat different, and so far, no one has determined the appropriate way to directly derive what the relaxation rate is. Using numerical data for the full solution, we extract the relaxation rate for an example and plot it in Fig.~\ref{fig:entropy_fig4}. One can see that is closely resembles the self-energy both in shape and magnitude, but also differs from it in important ways (particularly in not having the large peaks and in having more structure at low frequencies).

\begin{figure}
	\includegraphics[width=\columnwidth,clip=true,trim=0 0 0 0]{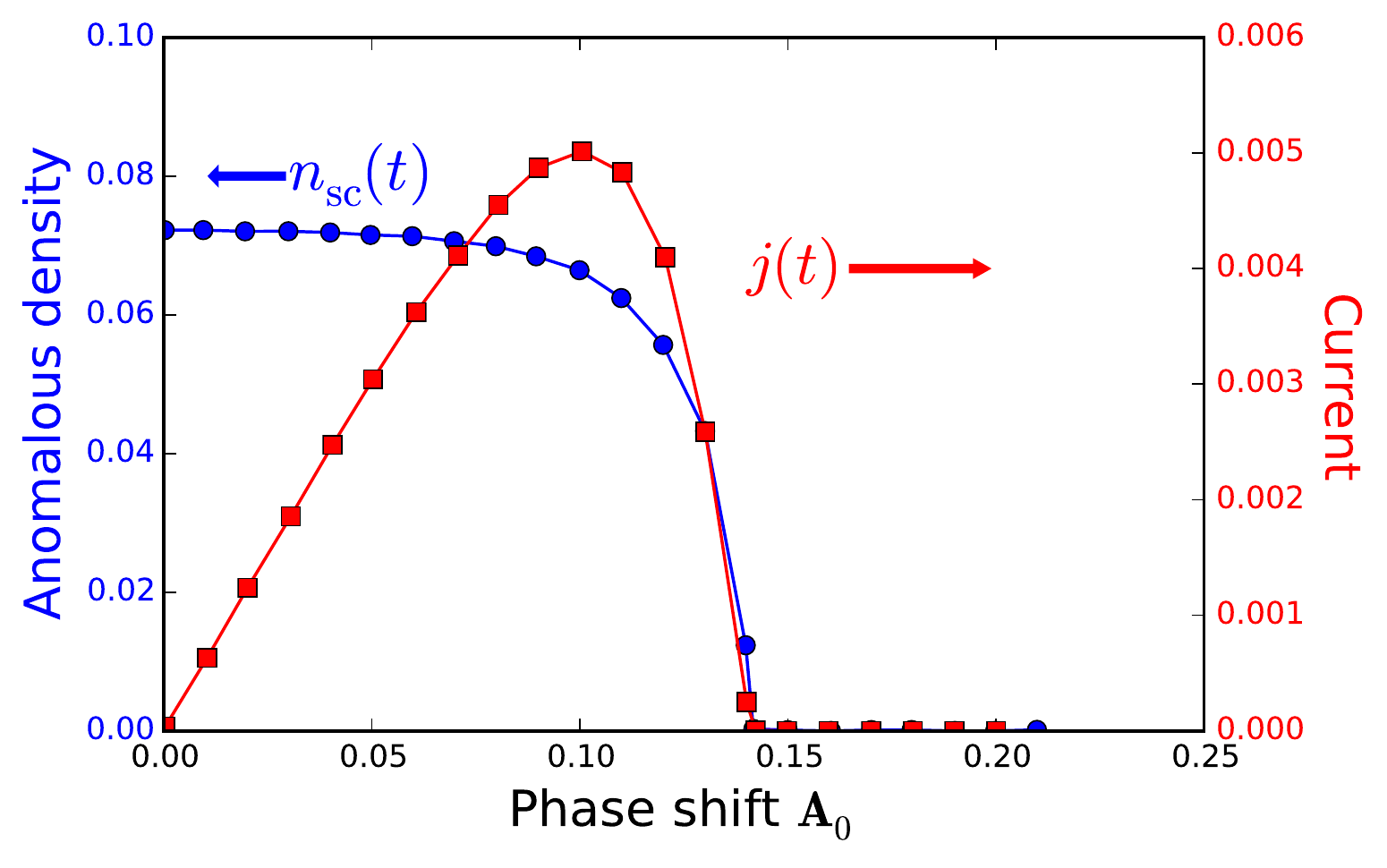}
	\caption{Superconductivity as a function of the constant phase shift $\A_0$ illustrating the
	supercurrent and the anomalous
	density $n_{sc}(t=0)$. }
	\label{fig:sc_delta_vs_A}
\end{figure}

\section{Time-resolved dynamics of the superconducting state}

Moving to the superconducting state, we now solve the equations of motions allowing for a finite
off-diagonal or ``anomalous'' component.  The equations of motion are otherwise equivalent,
with a $2 \times 2$ matrix in Nambu space representing the Green's function for each $t,t'$ on
the contour.

We first demonstrate that a superconducting solution is achieved by illustrating the
dependence of the solution and the current on the vector potential $\A(t)$. In the absence
of any superconductivity, there should be no effect of a time-independent $\A_0$, because that
vector potential corresponds to zero electric field. 
Fig.~\ref{fig:sc_delta_vs_A} shows a constant current proportional to the vector
potential as long as the anomalous density is not affected, and a drop in the supercurrent once
it does (we define the anomalous density as the analogue of the normal density,
$n_\mathrm{sc}(t)\equiv -i\sum_\kk F_\kk^<(t,t)$).

\begin{figure}[t]
	\includegraphics*[width=\linewidth,clip=true,trim=0 0 0 0]{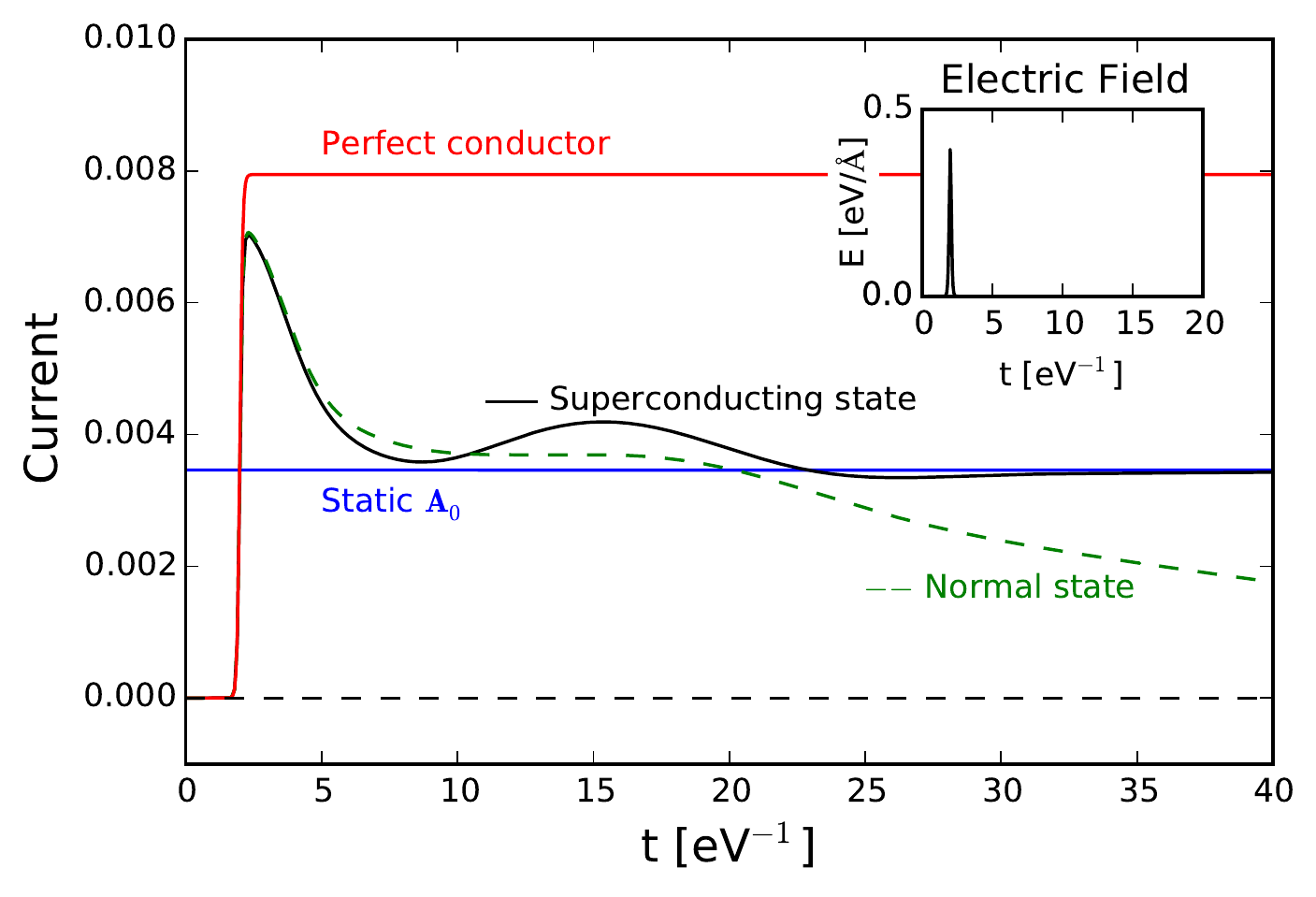}
	\caption{The resulting current due
	to a brief electric field pulse for the normal state, superconducting state, and a perfect
	metal.  For comparison, the superconductor with a constant static vector potential $\A_0$ is also
	shown.}
	\label{fig:supercurrents}
\end{figure}

Similarly, a brief applied electric field should lead to a supercurrent that remains present after
long times. Fig.~\ref{fig:supercurrents} shows the effect of a brief electric field pulse.  In the normal state,
a field pulse leads to a current that decays to 0 after some time. When superconductivity is present,
both a normal and a supercurrent are generated.  The normal current decays, leaving behind the
supercurrent. In fact, our results show that the remaining supercurrent is equal in magnitude to that
obtained from the approach above, where a constant vector potential shift is included.
Finally, in the absence of any interactions, the system is a \textit{perfect conductor}, and the
induced normal  current remains for all time.

\subsection{Higgs or amplitude oscillations}

\begin{figure*}[htpb]
	\includegraphics*[width=\textwidth,clip=true,trim=0 0 0 0]{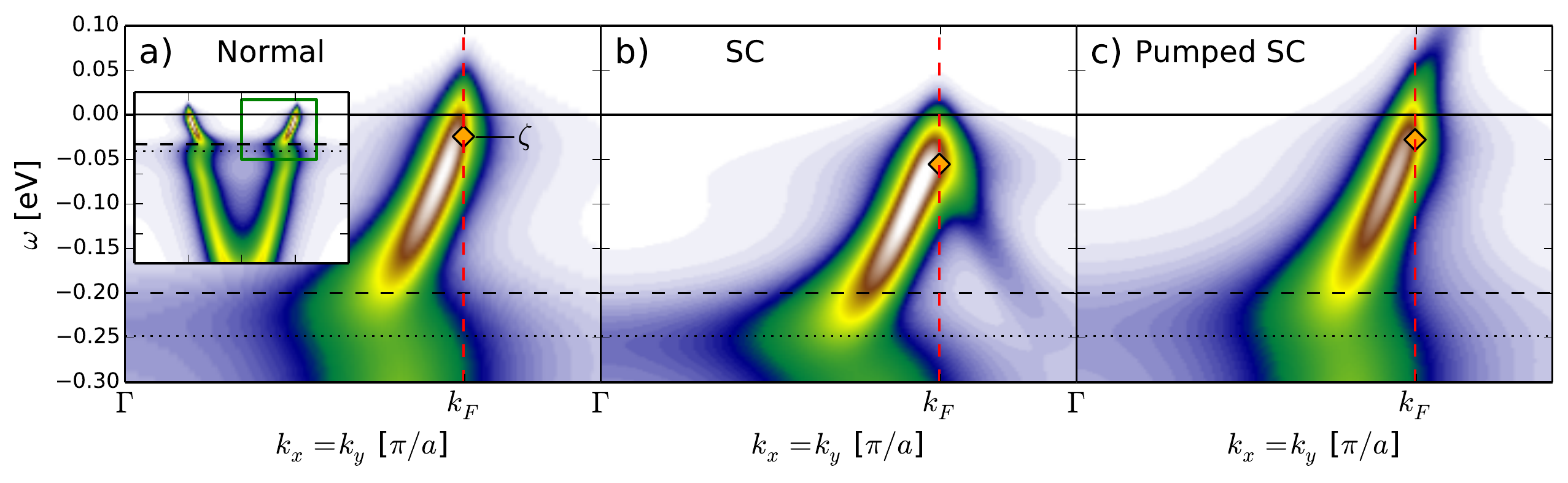}
	\caption{Time-resolved spectral functions near the Fermi level. Panels (a-c) show the
	normal state, the superconducting state, and a pumped superconductor, respectively.
	The inset illustrates the full spectrum in the normal state. Reprinted with permission from~\cite{kemper_direct_2015} \copyright ~American Physical Society.}
	\label{fig:higgs_spectra}
\end{figure*}

We next consider the single-electron spectra, shown in Fig.~\ref{fig:higgs_spectra}. Panel (a) 
shows the normal state spectrum for a coupled electron-phonon system, with the kink at 
$\Omega=0.2$ eV clearly visible. Panel (b) shows the spectrum in the superconducting state. The
spectral weight at the Fermi level has pulled back, indicating the opening of a gap. At the same
time, the kink has shifted down in energy by the magnitude of the gap $\Delta$, and a shadow band has appeared.
After applying a pump, as shown in panel (c), the spectrum looks more like the normal state
\textemdash the features that indicate the presence of superconductivity have all but disappeared.
\begin{figure*}[htpb]
\centering{
	\includegraphics*[width=0.9\textwidth,clip=true,trim=0 0 0 0]{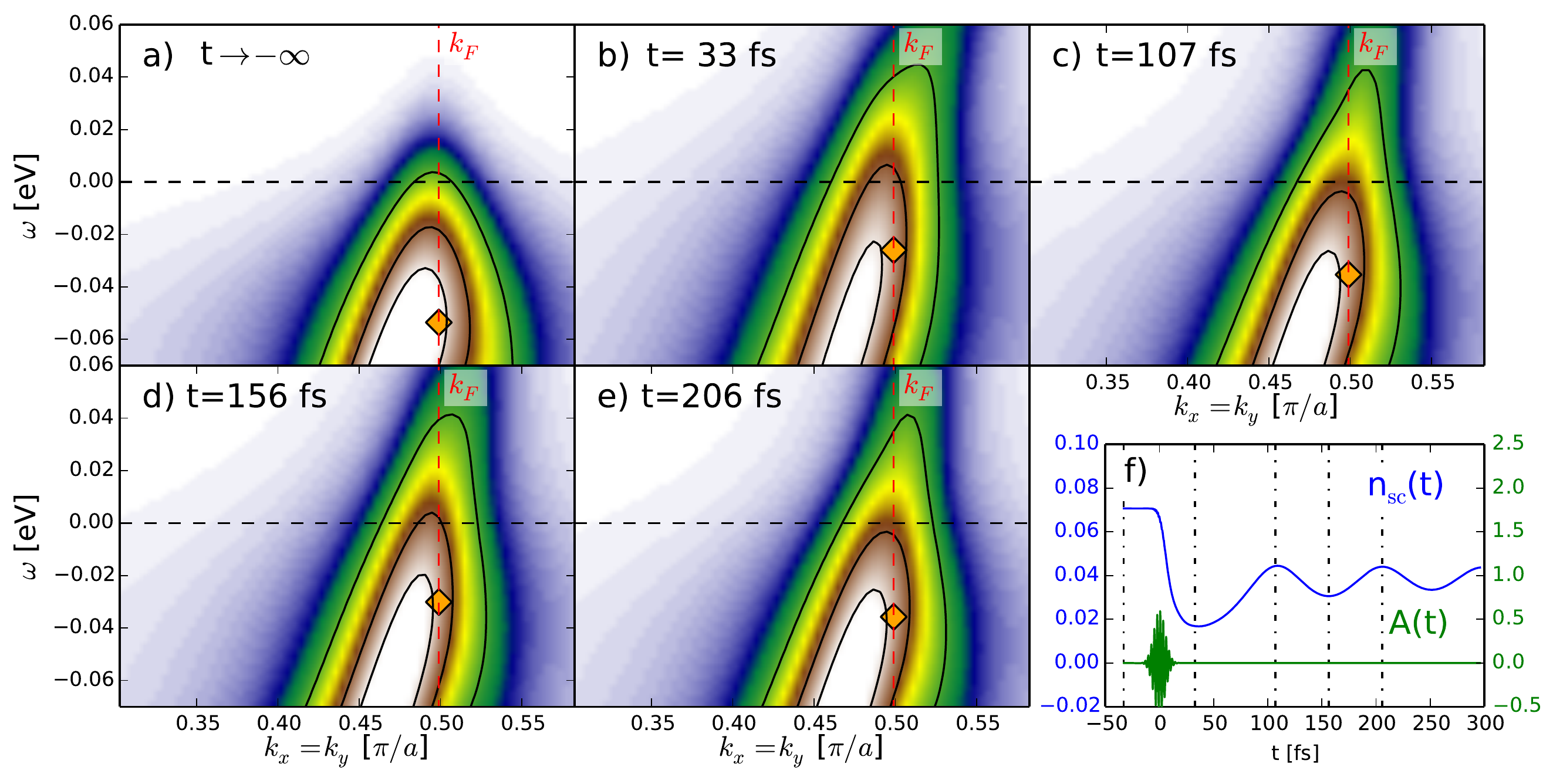}}
	\caption{Snapshots of the spectral function near $E_F$ illustrating the oscillations of the spectra as a function of time. The times are chosen to show the spectrum before the pump (a) and at the
	minima/maxima of the amplitude mode oscillations (b-e). Panel (f) shows the anomalous density
	as well as the applied vector potential $\A(t)$.  Reprinted with permission from\cite{kemper_direct_2015} \copyright ~American Physical Society.}
	\label{fig:higgs_snapshots}
\end{figure*}

However, the superconductivity has not entirely been eliminated. This can be observed by
considering the anomalous density 
which is plotted in 
Fig.~\ref{fig:higgs_snapshots}(f). The anomalous density is reduced, but remains finite after
the pump. The anomalous density also shows oscillations after the pump, which are reflected
in the snapshots of the tr-ARPES spectra (Figs.~\ref{fig:higgs_snapshots}b-e).
These oscillations occur at a frequency $\omega = 2 \Delta(\infty)$, where $\Delta(\infty)$ is the
remaining superconductivity after the pump\cite{volkov_74}. These are known as
Higgs oscillations, and arise from the amplitude mode of the superconductor. They have
been the subject of many studies using single-time BCS theory (see e.g.
\cite{volkov_74,kulik_81,yuzbashyan_2006,papenkort_2007,podolsky_2011,pekker_14,peronaci_transient_2015,krull_coupling_2016}),
but were discovered
here in a fully dynamical system.

\begin{figure}[htpb]
	\includegraphics[width=\linewidth,clip=true,trim=0 0 0 0]{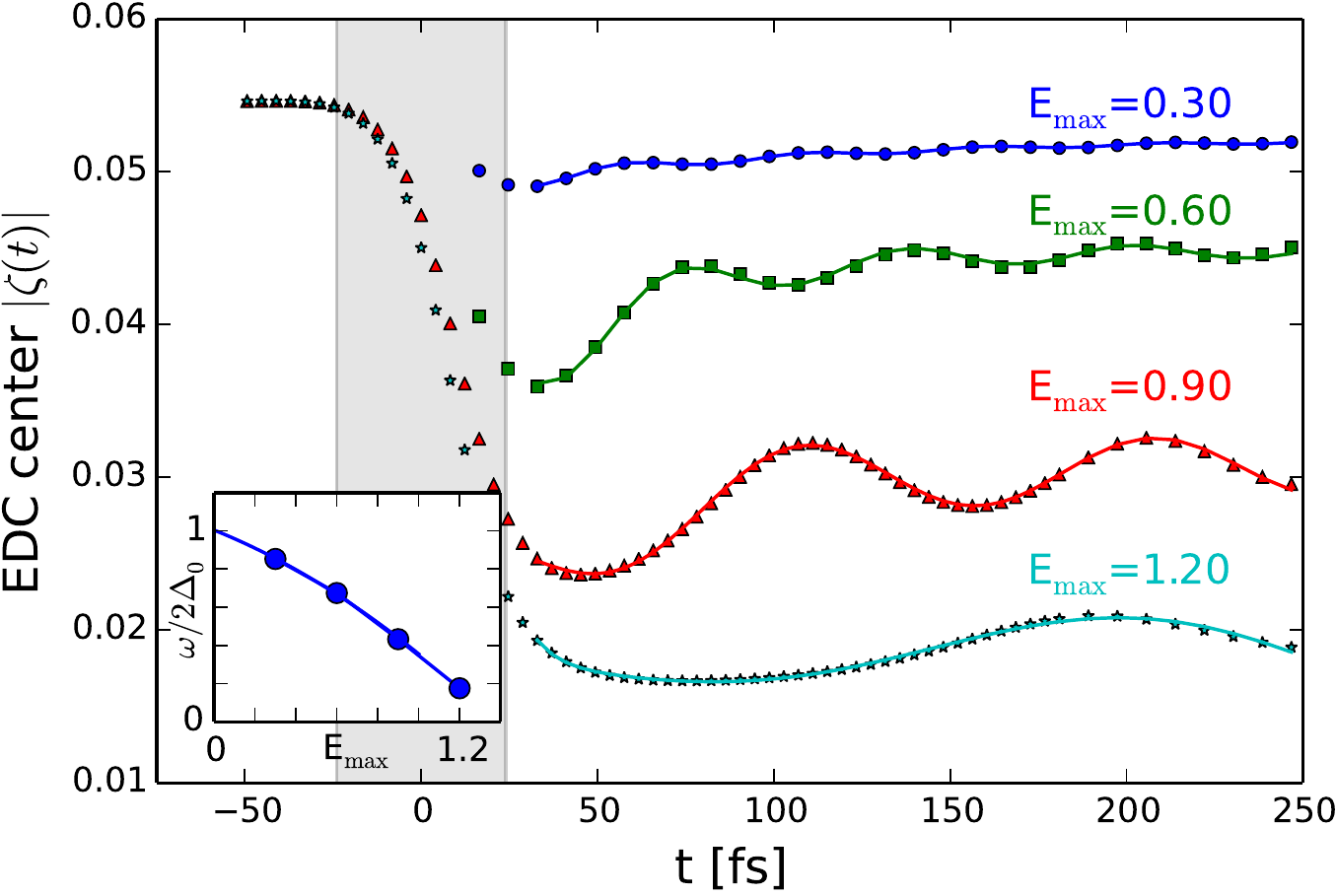}
	\caption{Oscillations in the EDC maximum as a function of time and pump
	fluence. Inset: normalized oscillation frequency vs. applied field. Reprinted with permission from\cite{kemper_direct_2015} \copyright ~American Physical Society.}
	\label{fig:higgs_vs_field}
\end{figure}

The oscillations are present throughout the spectrum, and can be obtained e.g. through an
analysis of the spectral weight above the Fermi level, or through the
position of the EDC maximum along some $\kk$\cite{kemper_direct_2015}. In either case,
we can perform a fluence dependence analysis. A lower fluence causes less melting of the SC
order, leading to a larger $\Delta(\infty)$ and faster oscillations.  Fig.~\ref{fig:higgs_vs_field} shows
the approach where the EDC maximum is analyzed and plotted as a function of time.  As expected,
the oscillations show a dependence on fluence (here represented by electric field amplitude), where
the oscillations speed up at lower fields. The tradeoff is that the amplitude also decreases.
From here, it can also be seen that it is critical that some superconductivity remains after the pump
to observe this phenomenon \textemdash if no SC is left, the oscillation frequency goes to 0. Hence, the effect is wiped out once the fluence becomes too high. These oscillations have been observed
recently using time-resolved THz transmission experiments\cite{matsunaga_14}
 but tr-ARPES has not yet
achieved a similar observation. This might be due to a difficulty in finding the proper fluence to see the Higgs oscillations.

\subsection{Light-enhanced superconductivity}
Recently, experiments have suggested that pump pulses can be used to enhance the
critical temperature of a superconductor in two quite different systems: the high-$T_C$ cuprates
and $K$-doped $C_{60}$\cite{fausti_light-induced_2011,mitrano_possible_2016}. The stated explanation for this effect lies in non-linear phononics, where
a resonantly excited phonon mode causes a non-oscillatory displacement in another mode.
The second, displaced mode causes some effect on either the electronic structure or the
pairing interactions, leading to an enhanced $T_C$\cite{mankowsky_nonlinear_2014}. 
Here we consider the former mechanism and model the change in the electronic structure by a
decrease of the hopping amplitude. This leads to an enhanced density of states at the Fermi level,
which in equilibrium would give an enhanced critical temperature\cite{schrieffer_handbook}.
We consider short and long ramp-down of the hopping interaction, and study the effect on the
system\cite{sentef_theory_2016}.
\begin{figure}[htpb]
	\includegraphics[width=\linewidth,clip=true,trim=0 0 0 0]{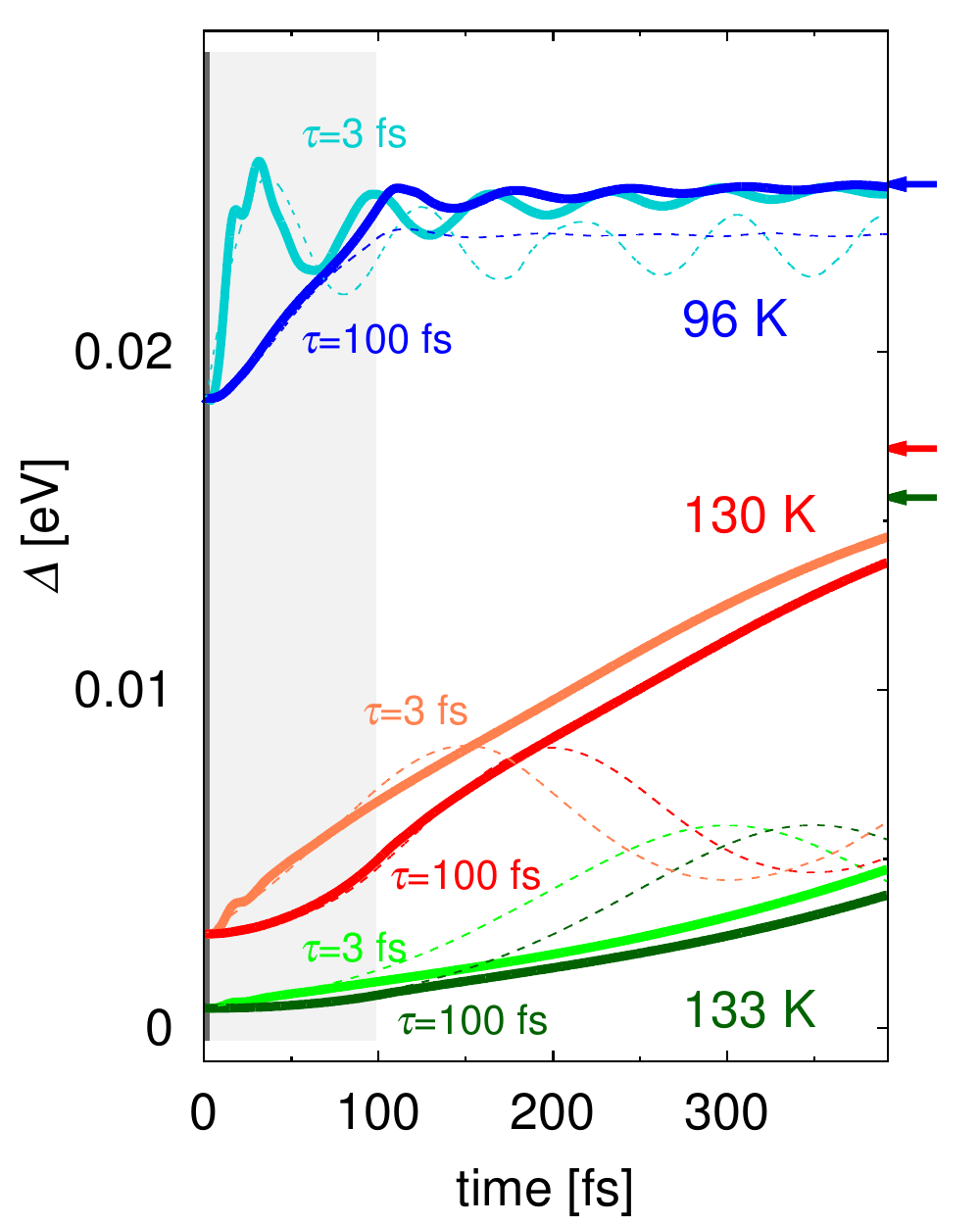}
	\caption{Order parameter $\Delta(t)$ during a short ramp (3 fs, light colors) and a long
	ramp (100 fs, dark colors). For comparison, the solution of a simple BCS model is
	shown with dashed lines. The equilibrium value of the gap for the final hopping
	parameter is shown as an arrow on the right side of the plot.
	Reprinted with permission from\cite{sentef_theory_2016} \copyright ~American Physical Society.}
	\label{fig:ramp_fig2}
\end{figure}

Fig.~\ref{fig:ramp_fig2} shows the effect of the changing bandwidth on the superconducting
gap $\Delta(t)$. For both long and short ramps, a marked increase in $\Delta(t)$ is observed,
with a much faster increase for the short ramp. After the ramps, $\Delta(t)$ continues to increase until
it reaches its equilibrium value, shown as arrows on the right side of the plot.  In some cases,
damped Higgs oscillations can also be observed. One particular observation is that the rate at
which $\Delta(t)$ increases depends linearly on the initial $\Delta(t=0)$.
The figure also shows a comparison to a Bardeen-Cooper-Schrieffer (BCS) model with equivalent $\Delta(t=0)$ and
$\Delta(t=\infty)$. The BCS model agrees mainly at short times, when the overall
dynamics are captured by the changes in the coherence factors. The long-time behavior, where
dissipation starts to play an important role is not captured by BCS.

\section{Summary}

In this work, we have summarized a series of papers which shed light onto tr-ARPES experiments using a pump/probe excitation and detection scheme. We found that the thermalization of the combined electron-phonon system is complex and not simply governed by quasiequilibrium relaxation rates. We also showed how spectral weight distributions affect the ``phonon window effect'' and the kink feature in the normal state. Finally, we examined a number of features in the superconducting state, including Higgs (amplitude-mode) oscillations and light-enhanced superconductivity. This work has only touched the tip of the iceberg in determining the behavior of these complex systems. Major open problems include questions such as the following: (i) how does the long-time relaxation change when the finite heat capacity of the phonons is taken into account? (ii) what happens to the self-energy sum rules in the superconducting state? (iii) what effect does order parameter symmetry have on the nonequilibrium properties of a superconductor? (iv) how can one determine microscopic relaxation rates from a semi-analytic theory? and (v) what do these nonequilibrium measurements tell us about equilibrium?

We hope that the field will engage in these open questions and work on answering them in the near future. We also look forward to more surprises coming from experiment which will need increasingly more detailed and materials-specific theory to be able to describe the phenomena.

\section{Acknowledgments}

This work was supported by the Department of Energy, Office of Basic Energy Sciences, Division of Materials Sciences and Engineering under Contracts No. DE-AC02-76SF00515 (Stanford/SIMES), and No. DE-FG02-08ER46542 (Georgetown).
Computational resources were provided by the National Energy Research Scientific Computing Center supported by the Department of Energy, Office of Science, under Contract No. DE-AC02-05CH11231. 
J.K.F. was also supported by the McDevitt Bequest at Georgetown.

\bibliographystyle{andp2012}
\bibliography{tdrefs}

\end{document}